\documentclass[lettersize,journal]{IEEEtran}
\usepackage{amsmath,amsfonts,amssymb,bm}
\usepackage{algorithm}
\usepackage{algpseudocode}
\usepackage{array}
\usepackage[caption=false,font=normalsize,labelfont=sf,textfont=sf]{subfig}
\usepackage{textcomp,xcolor}
\usepackage{stfloats}
\usepackage{url}
\usepackage{verbatim}
\usepackage{graphicx}
\usepackage{cite}
\usepackage{multirow}
\usepackage{rotating}
\usepackage{makecell}
\usepackage{tabularx}

\newcolumntype{C}[1]{>{\centering\arraybackslash}m{#1}}
\newcolumntype{Y}{>{\centering\arraybackslash}X}

\usepackage{times}
\usepackage{epsfig}
\usepackage{lipsum}  
\usepackage{soul}
\usepackage{subcaption}

\begin{document}

\setlength{\textfloatsep}{4pt}
\setlength{\intextsep}{4pt}
\setlength{\floatsep}{4pt}

\title{Securing Distributed Network Digital Twin Systems Against Model Poisoning Attacks}

\author{
		\IEEEauthorblockN{Zifan Zhang,
            Minghong Fang,
            Mingzhe Chen,~\IEEEmembership{Member,~IEEE,} \\
            Gaolei Li,~\IEEEmembership{Member,~IEEE,} 
		  Xi Lin,~\IEEEmembership{Member,~IEEE,}
            Yuchen Liu,~\IEEEmembership{Member,~IEEE} \\
 }

\thanks{This work was supported by the National Science Foundation under Award SaTC--2350075, CNS--2312138, CNS--2312139, and SaTC--2350076.}
\thanks{
Z. Zhang and Y. Liu are with the Department of Computer Science, North Carolina State University, Raleigh, NC, 27695, USA (Email: \{zzhang66, yuchen.liu\}@ncsu.edu).
\textit{(Corresponding author: Yuchen Liu.)}}
\thanks{Minghong Fang is with the Department of Computer Science and Engineering, University of Louisville, Louisville, KY, 40208, USA (Email: \protect\url{ minghong.fang@louisville.edu)}.}
\thanks{M. Chen is with the Department of Electrical and Computer Engineering and Frost Institute for Data Science and Computing, University of Miami, Coral Gables, FL 33146 USA (Email: \protect\url{ mingzhe.chen@miami.edu)}.} 
\thanks{G. Li and X. Lin are with the School of Electronic Information and Electrical Engineering
Shanghai Jiao Tong University, China (Email: \{gaolei\_li, linxi234\}@sjtu.edu.cn).}
\vspace{-1.5em}
}
\markboth{Internet of Things Journal}%
{Shell \MakeLowercase{\textit{et al.}}: A Sample Article Using IEEEtran.cls for IEEE Journals}


\maketitle

\begin{abstract}

In the era of 5G and beyond, the increasing complexity of wireless networks necessitates innovative frameworks for efficient management and deployment. Digital twins (DTs), embodying real-time monitoring, predictive configurations, and enhanced decision-making capabilities, stand out as a promising solution in this context. 
Within a time-series data-driven framework that effectively maps wireless networks into digital counterparts, encapsulated by integrated vertical and horizontal twinning phases, 
this study investigates the security challenges in distributed network DT systems, which potentially undermine the reliability of subsequent network applications such as wireless traffic forecasting.
Specifically, we consider a minimal-knowledge scenario for all attackers, in that they do not have access to network data and other specialized knowledge, yet can interact with previous iterations of server-level models. In this context, we spotlight a novel fake traffic injection attack designed to compromise a distributed network DT system for wireless traffic prediction. In response, we then propose a defense mechanism, termed global-local inconsistency detection (GLID), to counteract various model poisoning threats. GLID strategically removes abnormal model parameters that deviate beyond a particular percentile range, thereby fortifying the security of network twinning process. Through extensive experiments on real-world wireless traffic datasets, our experimental evaluations show that both our attack and defense strategies significantly outperform existing baselines, highlighting the importance of security measures in the design and implementation of DTs for 5G and beyond network systems. 
\end{abstract}


\section{Introduction}
\label{sec:intro}

In the realm of telecommunications, wireless networks are experiencing a paradigm shift, primarily driven by the advent of edge computing, spectrum sharing, and millimeter-wave communication technologies in the 5G era. These technological advancements are foundational to a multitude of novel applications and services, notably enhancing mobile broadband and facilitating the seamless integration of the Internet of Things (IoT)~\cite{nguyen20216g, qadir2023towards}, autonomous transportation~\cite{miglani2019deep}, smart urban infrastructure~\cite{cugurullo2020urban}, and remote healthcare delivery~\cite{qayyum2020secure}. Further, the nascent stages of 6G research are indicative of potential revolutionary leaps in hybrid physical-virtual network technologies, paving the way for ubiquitous and intelligent connectivity worldwide.
Parallel to these advancements, the concept of digital twin (DT) has surfaced as a significant technological breakthrough in the mixed reality era~\cite{vanderhorn2021digital, semeraro2021digital, botin2022digital}. The DTs embody intricate virtual representations of physical entities or systems and gain traction in the context of the Fourth Industrial Revolution. This concept synergistically harnesses the capabilities of IoT, machine learning, and big data analytics, meticulously constructing a comprehensive digital model that mirrors the physical attributes, processes, interconnection, and dynamics of its real-world counterpart. Such models play a pivotal role in facilitating predictive simulations, what-if analysis, and system optimizations within a virtual environment, thereby offering tangible insights into operational challenges and maintenance requirements~\cite{feng2023digital, yu2022energy,li2022digital}.

While DTs offer a wide range of benefits and applications, ensuring their security remains a critical concern that necessitates a comprehensive understanding and robust countermeasures. Common security threats to DTs include data breaches, unauthorized access, and cyber-attacks, which can disrupt the seamless interaction between the physical and virtual systems~\cite{alcaraz2022digital}. In the realm of wireless networks, these DTs face additional challenges such as Byzantine attacks, man-in-the-middle attacks, and signal interference, which can severely impact their availability and reliability. Furthermore, robust countermeasures, including encryption techniques and secure communication protocols, are needed to protect DTs from potential adversarial attacks in open wireless environments~\cite{karaarslan2021digital}. As DTs are increasingly integrated into the metaverse applications~\cite{ning2023survey}, addressing the security and privacy challenges becomes paramount to ensure a trustworthy user interface~\cite{far2022applying}. In particular, trust evaluation schemes such as in~\cite{10234616} have been proposed for using federated learning (FL) in DT systems, aiming to enhance data usage security by evaluating the trustworthiness of participating network entities.
In the realm of wireless networks, FL leverages its decentralized nature to facilitate multiple network services. With the exponential growth in the number of connected devices and the ever-increasing demand for data-intensive applications like streaming and IoT services, constructing precise network digital twins (NDTs) accurately becomes vital for ensuring various downstream forecasting tasks, 
such as wireless traffic prediction (WTP)~\cite{qiu2018spatio, xu2017high, zhang2021dual}. 
Despite distributed learning's potential in accuracy, efficiency, and privacy preservation, its integration into NDT creation and operation is not devoid of challenges. Notably, Byzantine attacks, particularly model poisoning attacks, pose significant threats to the effectiveness and trustworthiness of NDT systems. 

In a model poisoning attack, malicious network entities introduce adversarial modifications to the model parameters during the mapping process of NDTs. This tampering results in a compromised server-level twin, i.e. global twin model, when aggregated at the central network controller, subsequently producing incorrect operations on the physical infrastructure. Such inaccuracies lead to the risk of network inefficiencies and even severe service disruptions, especially in real-time applications like autonomous driving systems. In more extreme scenarios, these attacks may serve as gateways to further malicious network intrusions, instigating broader security and privacy concerns as illustrated in~\cite{joshi2015review, fan2019research}. The grave implications of model poisoning attacks underscore the pressing need for robust security measures to ensure the integrity, reliability, and resilience of distributed NDT systems against Byzantine failures, thereby safeguarding the overarching network infrastructure and the services reliant on it. 
While most existing DT mapping algorithms and their associated security strategies are typically assessed within the context of \textit{classification} problems~\cite{fang2020local,shejwalkar2021manipulating}, scant attention has been paid to the \textit{regression} problems, as observed in examined WTP scenarios within NDTs, introducing distinct challenges related to data distribution, model complexity, and evaluation metrics. The distinction between data manipulation strategies in regression and classification problems, as well as their detection methodologies, underscores the nuanced challenges in safeguarding twin models against emerging adversarial attacks. For instance, in a regression-based DT-assisted WTP problem, attackers typically target the model's continuous output by altering the distribution or magnitude of input time-series data, intending to steer predictions in a specific direction. This differs from classification tasks, where the manipulation revolves around modifying input features to induce misclassification without noticeably changing the input's appearance to human observers.

To bridge this gap, we make the first attempt to introduce a novel attack centered on injecting disruptive traffic data from 
malicious NDTs into wireless networks. Existing model poisoning attacks have predominantly depended on additional access knowledge and direct intrusions on physical base stations (BSs)~\cite{zheng2022poisoning, fang2020local, xie2020fall}. However, in a practical cellular network system, BSs have exhibited a commendable level of resilience against attacks, making the extraction of training data from them a challenging endeavor. In contrast, the cost of deploying fake NDTs that mimic their behaviors is comparatively lower than the resources required for compromising authentic BSs~\cite{cao2022mpaf}. This assumption asserts that these compromised NDTs lack insight into the training data and only have access to the initial and current global twin models, aligning with the practical settings studied in~\cite{cao2022mpaf}. 
Importantly, other information, such as data aggregation rules and model parameters from benign NDTs or BSs, remains inaccessible to these compromised NDTs. 
In this work, we consider a distributed DT-assisted network architecture as depicted in Fig.~\ref{fig:mapping}, where wireless traffic data collected from BSs is mapped into local NDTs to establish an initial and private NDT for each BS. Within each cluster, a cluster-level NDT (C-NDT) is constructed by aggregating these local twin models.
Subsequently, at the backend, a global twin model (G-NDT) is established by merging the C-NDT model parameters during each iterative phase. This global twin model is then synchronized with each local NDT, serving as a foundation for predictive analysis and enabling specific applications for each BS and its associated NDTs.
In this situation, our threat model envisions a minimum-knowledge scenario for an adversary. First, we propose Fake Traffic Injection (FTI), a methodology designed to create undetectable fake NDTs with minimal prior knowledge. Each fake NDT employs both its initial model and current global information to determine the optimizing trajectory of the twinning process, as shown in Fig.~\ref{fig:security}. These malicious participants aim to subtly align the global model towards an outcome that undermines the integrity and reliability of the NDT system. Numerous numerical experiments are conducted to validate that our FTI demonstrates efficacy across various state-of-the-art model aggregation rules, outperforming other poisoning attacks in terms of vulnerability impacts.

On the contrary, we propose an innovative defensive strategy known as Global-Local Inconsistency Detection (GLID), aimed at neutralizing the effects of model poisoning attacks on NDT systems. This defense scheme involves strategically removing abnormal model parameters that deviate beyond a specific percentile range estimated through statistical methods in each dimension. Such an adaptive approach allows us to trim varying numbers of malicious model parameters instead of a fixed quantity~\cite{yin2018byzantine}. Next, a weighted mean mechanism is employed to update the global twin model parameter, subsequently disseminated back to each NDT. Our extensive evaluations, conducted on real-world datasets, demonstrate that the proposed defensive mechanism substantially mitigates the impact of model poisoning attacks on NDT systems, thereby showcasing a promising avenue for securing distributed NDT systems with trustworthiness. 
This paper is an extended version of our previous work in~\cite{zhang2024poisoning}, where we expand upon it by adapting the proposed attack and defense strategies from traditional federated learning settings to the practical NDT system.

The contributions are briefly summarized into three folds:
\begin{enumerate}
    \item We present a novel model poisoning attack, employing fake NDTs for traffic injection into distributed NDT systems under a minimum-knowledge scenario.
    \item Conversely, we propose an effective defense strategy tailored to counteract various model poisoning attacks, which proactively trims an adaptive number of twin model parameters by leveraging the percentile estimation technique.
    \item Lastly, we evaluate both the proposed poisoning attack and the defensive mechanism using real-world traffic datasets from Milan City, where the results demonstrate that the FTI attack indeed compromises distributed NDT systems, and the proposed defensive strategy proves notably more effective than other baseline approaches in mitigating various attacks.
\end{enumerate}

\section{Related Works and Preliminaries}
\label{sec:related}

\subsection{Distributed Network Digital Twin Systems}

The integration of DTs into the realm of wireless networking represents a significant leap forward in this rapidly evolving field. As outlined in~\cite{networkdigitaltwin}, the use of DTs involves the creation of detailed virtual replicas of network components and infrastructure. This approach enables real-time analytics and optimization, providing deep insights into network behavior under various scenarios. Such strategies are crucial for predictive maintenance and performance monitoring, greatly enhancing network reliability and efficiency. 
Furthermore, \cite{graphneuralnetwork} introduces the use of Graph Neural Networks to enhance DTs in network slicing, aimed at predicting network performance and optimizing resources in high-bandwidth and low-latency scenarios. Additionally, the application of DTs in vehicular networks is detailed in \cite{intelligentdigitaltwin}, showcasing DTs' ability to model and control software-defined vehicular networks, thereby improving the effectiveness and reliability of vehicular communications. 
Moreover, \cite{10244089} proposes a digital twin-assisted security scheme for multi-resource heterogeneous RANs in space-air-ground integrated networks.
Despite various explorations into DT applications within wireless networks, there is a gap in the literature regarding the development and mapping of NDTs, which our research seeks to address.

At the forefront of DTs, distributed learning emerges as a revolutionary approach, especially for large-scale networks and the Industrial IoT. The combination of these technologies not only enhances system efficiency but also transforms data handling capabilities. 
\cite{zhang2024mapping} proposed a Joint Vertical and Horizontal Learning-based digital twinning strategy to perform a precise mapping from physical networks to digital twins.
\cite{zhang2024REC} exemplifies their potential in reliable edge caching and real-time data-driven optimization.
Additionally, addressing the challenge of efficient data communication in distributed learning systems, \cite{communication_efficient_federated} and \cite{communication_efficient_iot} propose strategies to enhance data exchange and processing—crucial for scaling up applications with massive access. 
\vspace{-0.4cm}
\subsection{Poisoning Attacks on Distributed Systems}

The decentralized architecture of distributed DT systems renders them vulnerable to Byzantine attacks, as explored in previous study~\cite{fang2020local,cao2022mpaf,shejwalkar2021manipulating,zheng2022poisoning,tolpegin2020data,10459057}. In these scenarios, adversaries can compromise BSs and their corresponding NDT models to undermine the entire distributed DT system. These malicious BSs may tamper with their local training data or directly modify their local twin models to negatively impact the global twin model.
For example, the Trim attack~\cite{fang2020local} involves malicious BSs deliberately distorting their local twin models to create a significant discrepancy in the aggregated model post-attack compared to its pre-attack state.
The MPAF attack~\cite{cao2022mpaf} sees each compromised BS applying a negative scalar to the global twin model update before forwarding this tampered update to the server-level twin.
In the Random attack~\cite{fang2020local}, malicious DTs generate and send a random vector, drawn from a Gaussian distribution, to the server as their update.
Furthermore, a recent study by \cite{zheng2022poisoning} introduced specific poisoning attacks targeting distributed DT systems. Here, an attacker manipulates some BSs under their control, each with its local training dataset. These DTs adjust their local models using this data and then scale their model updates by a factor before dispatching these altered updates to the server. These strategies highlight the critical challenge of securing the entire system against various forms of data and model tampering attacks, underscoring the need for robust defense mechanisms.

Existing attacks in our considered setting suffer from the following limitations. In the MPAF attack, model updates from fake NDTs are exaggerated by a factor such as $1 \times 10^6$. However, this approach is impractical because the server can easily identify these excessive updates as anomalies and discard them. Furthermore, such blatant manipulation lacks subtlety, making it easy to detect and counter. On the other hand, our method involves carefully crafting model updates on fake clients by solving an optimization problem. This ensures that the server is unable to differentiate these fake updates from benign ones, allowing the attacker to simultaneously breach the integrity of the system without detection. Our approach maintains the updates within a plausible range, avoiding the pitfalls of easily detectable anomalies.
The attack described in~\cite{zheng2022poisoning} is not feasible because it is based on the unrealistic assumption that an attacker can easily take control of authentic BSs or DTs. In reality, it is highly challenging for an attacker to gain such influence over existing, authentic facilities. Moreover, this attack does not consider the sophisticated security measures typically in place to protect these systems. Our research, however, focuses on developing more realistic attack scenarios that account for the complexities and security protocols of modern distributed DT systems, ensuring a more accurate assessment of their vulnerabilities.

\vspace{-0.4cm}
\subsection{Byzantine-robust Aggregation Rules}

In environments free from adversarial intentions, server-level twins typically aggregate incoming local twin model updates through a simple averaging process~\cite{mcmahan2017communication}. However, recent studies~\cite{blanchard2017machine, zhang2024poisoning} have revealed vulnerabilities in this averaging method of aggregation, particularly its susceptibility to poisoning attacks. In such attacks, a single malicious local twin model can significantly alter the aggregated result. To counter these vulnerabilities, the literature offers a range of Byzantine-resistant aggregation algorithms~\cite{blanchard2017machine,yin2018byzantine,cao2020fltrust,fung2018mitigating,sharma2023flair,xia2019faba, fang2024DFL}.
For instance, the Krum method~\cite{blanchard2017machine} assesses each local twin's update by calculating the sum of Euclidean distances to updates from other twin models, selecting the update with the smallest sum for global aggregation.
The Median aggregation strategy~\cite{yin2018byzantine} involves the server-level twin computing median values across each parameter from all local updates, improving resistance to outlier manipulations.
These strategies introduce robustness against adversarial actions, ensuring the integrity of the aggregated twin model in distributed systems.
\section{Creation and Synchronization of Network Digital Twins for Wireless Traffic Prediction} 
\label{sec:dt}

\begin{figure*}
    \centering
    \includegraphics[scale = 0.35]{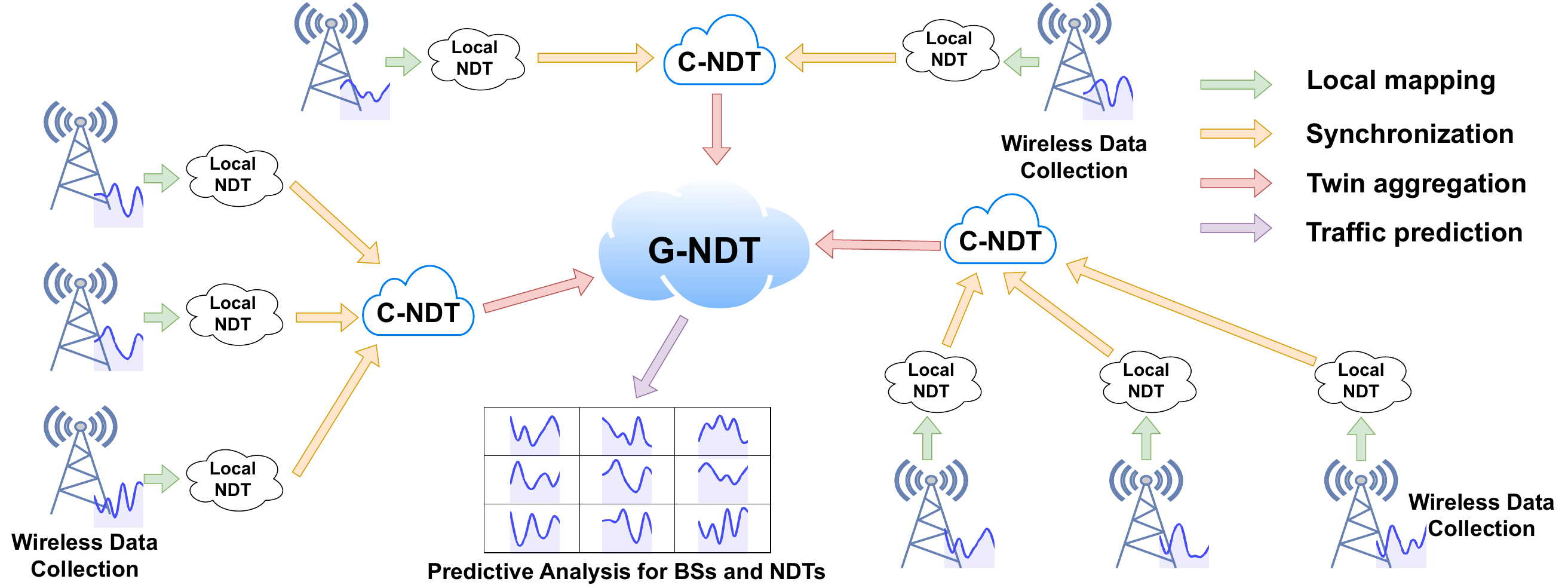}
    \caption{Distributed network digital twin framework for wireless traffic prediction.}
    \vspace{-0.2in}
    \label{fig:mapping}
\end{figure*}

This section introduces a novel framework for creating and synchronizing NDTs specifically designed for wireless traffic prediction. The framework is structured around three main stages: dynamic connectivity segmentation (DCS), vertical twinning (V-twinning), and horizontal twinning (H-twinning). The overall framework is shown in Fig.~\ref{fig:mapping}.
The primary objective of the NDTs is to minimize prediction errors across all BSs for a better understanding of the physical network. This can be formulated as an optimization problem:
\begin{equation}
\label{all_obj}
    \bm{\alpha}^* = \arg\min_{\bm{\alpha}}  \frac{1}{Mz} \sum_{m=1}^{M} \sum_{n=1}^{z} F(f(r_m^n, \bm{\alpha}), s_m^n),
\end{equation}
where $F$ is the quadratic loss function, $M$ is the number of NDTs, $z$ is the number of data points, $r_m^n$ is the input traffic sequence, and $s_m^n$ is the corresponding output traffic prediction. The optimization problem is resolved through FL with distributed NDTs, following the synchronization, local updating, and model aggregation process. This single-level mapping approach utilizes a classical FL strategy to aggregate multiple local twin models, serving as the baseline for comparing with our proposed joint vertical-horizontal mapping scheme.

\begin{figure}[ht]
    \centering
    \includegraphics[scale = 0.54]{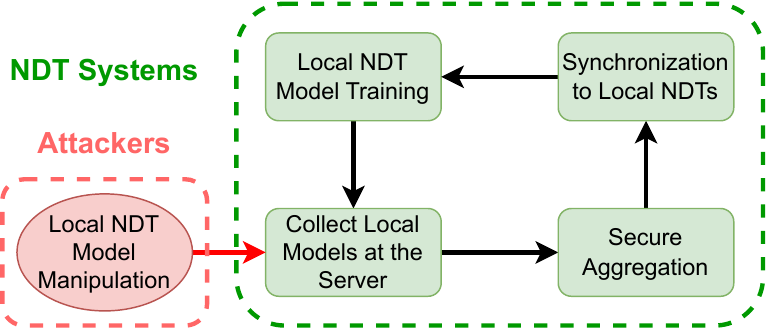}
    \caption{Process of model poisoning attack and secure aggregation defense.}
    \label{fig:flowchart}
\end{figure}

Specifically, Eq.~(\ref{all_obj}) can be resolved in a distributed fashion in traditional FL settings with the following three steps in each global training round $t$, as shown in Fig.~\ref{fig:flowchart}.

\begin{itemize}

    \item \textbf{Step I (Local twin update).}
    Each NDT $i \in [n]$ utilizes its private time-series training data along with the current global model to refine its own local model, then transmits the updated local model $\bm{\theta}_i^{t}$ back to a central server.

    \item \textbf{Step II (Local twin manipulation/model poisoning attack).}
    Each malicious NDT utilizes its knowledge to modify or create local twin models, and then send these malicious twin models to the server.

    \item \textbf{Step III (Aggregation of local twin models).}
    The central server leverages the aggregation rule ($\text{AR}$) to merge the $n$ received local models and subsequently updates the global model as follows:
    \begin{align}
    \bm{\theta}^{t+1} = \text{AR}\{\bm{\theta}_1^{t}, \bm{\theta}_2^{t}, \ldots, \bm{\theta}_n^{t}\}.
    \end{align}
     The commonly used aggregation rule is the FedAvg~\cite{mcmahan2017communication}, where the server simply averages the received $n$ local models from distributed NDTs, i.e., $\text{AR}\{\bm{\theta}_1^{t}, \bm{\theta}_2^{t}, \ldots, \bm{\theta}_n^{t}\} = \frac{1}{n}\sum\limits_{i=1}^n \bm{\theta}_i^{t}$.

     \item \textbf{Step IV (Synchronization).}
    The central server sends the current global model $\bm{\theta}^t$ to all NDTs.
\end{itemize}

Specifically, our multi-level mapping framework encompasses a central DT, named global network digital twin (G-NDT), coordinating with a network of \( M \) NDTs, and multiple cluster network digital twin (C-NDT). Each NDT, denoted as \( m \) in the set \([M]\), independently holds a proprietary dataset \( d_m = \{d^1_m, d^2_m, \ldots, d^L_m\} \). In this dataset, \( L \) indicates the total number of time intervals, and \( d^l_m \) represents the traffic load at NDT \( m \) during the \( l \)-th interval, where \( l \) ranges over \([L]\). The NDT involves constructing input-output predictive traffic sequences locally, denoted as \( \{r_m^n, s_m^n\}_{n=1}^{z} \), for each NDT to generate future traffic predictions. Here, \( r_m^n \) is a suNDTet of historical traffic data corresponding to the output \( s_m^n =\{d_m^{l-1}, \dots, d_m^{l-a}, d_m^{l-\rho 1},\dots, d_m^{l-\rho b}\} \). The parameters \( a \) and \( b \) represent sliding windows that capture immediate and cyclical temporal dependencies, respectively, while \( \rho \) reflects inherent periodicities in the network, which might be influenced by user activity patterns or application service demands.

The first stage, DCS, is employed periodically to ensure effective clustering of NDTs with similar communication characteristics and networking configurations. This clustering step is integral to the efficient creation and updates of multiple distributed NDTs, i.e. C-NDTs, which demonstrate distinct behaviors and perform parallel synchronization with the G-NDT. The DCS algorithm clusters the NDTs based on attributes such as geological distances, capacity of backhaul links, coverage area overlaps, and similarity of frequency of occurrence distribution. The relationship between two NDTs, $n_1$ and $n_2$, is quantified by a metric $\Phi_{n_1,n_2}$:

\begin{equation}
    \Phi_{n_1,n_2} = \frac{\omega_g}{g_{n_1,n_2}} + \omega_k \cdot k_{n_1,n_2} + \omega_\beta \cdot \beta_{n_1,n_2} + \omega_\tau \cdot \tau_{n_1,n_2},
\end{equation}

\noindent where $\omega$ represents the weights for each attribute. This dynamic clustering enhances the twinning performance in real time and forms the basis for accurate wireless traffic prediction by grouping NDTs with similar traffic patterns.

In the V-Twinning stage, initial NDTs are created with historical data on caching requests and their frequency. It employs an FL strategy, where model parameters are shared among NDTs instead of raw data, enabling collaborative training of a global model. This approach efficiently distributes twinning tasks across NDTs while ensuring content data privacy. Specifically, the V-Twinning stage initializes a concrete G-NDT and synchronizes C-NDTs with the G-NDT after the twinning aggregation process. The aggregation of C-NDTs to form the G-NDT is given by:
\begin{equation}
    \bm{\alpha}^{t+1} = \frac{1}{C}\sum_{c=1}^C \bm{\alpha}_c^t,
\end{equation}
where $\bm{\alpha}_c^t$ represents the model parameters of the C-NDT for cluster $c$ at time $t$, and $C$ is the number of clusters. This stage is crucial for initializing the network DTs with historical traffic data, which serves as a foundation for future traffic prediction.

The H-Twinning stage is designed to periodically synchronize the physical network and NDTs with real-time data. It adopts an asynchronous FL approach to update with dynamics from the physical network, providing a scalable and flexible solution for wireless networks composed of multiple clusters. This stage updates the twins regularly, ensuring that all NDTs remain relevant and accurately simulate and predict wireless traffic patterns. The update rule for the G-NDT based on the deviation $\epsilon$ between a C-NDT and the current G-NDT is as follows:
\begin{equation}
    \bm{\alpha}^{t+1} = 
    \begin{cases} 
    \frac{1}{C}\sum_{c=1}^C \bm{\alpha}_c^t & \text{if } \epsilon > \psi, \\
    \bm{\alpha}^t & \text{otherwise},
    \end{cases}
\end{equation}
where $\psi$ is a predefined threshold, and $\epsilon = (\bm{\alpha}_c^t - \bm{\alpha}^t)^2$ measures the deviation between the C-NDT and the G-NDT. This stage is critical for incorporating real-time traffic data into the NDTs, enabling them to adapt to changing network conditions and improve traffic prediction accuracy.

\section{Threat Model for Distributed Network Digital Twins}
\label{sec:threat}

Built upon the constructed distributed NDT system, this section discusses the threat model and explores a novel attack that poses a security breach to system functionality and network operations. 
\vspace{-0.15in}
\subsection{Objective of the Attacker}
The fundamental aim of an attacker targeting a distributed NDT system is to impair the performance of the composite global twin model significantly. Such impairment directly undermines the precision of real-time traffic forecasts, which is crucial for effective network management and resource distribution. The ramifications of compromised traffic predictions include network congestion, diminished service quality, and suboptimal resource utilization, presenting considerable operational hurdles for network operators. This disturbance extends beyond the service providers, affecting end-users dependent on stable and efficient network services.
\vspace{-0.15in}
\subsection{Capabilities of the Attacker}
\begin{figure}
    \centering
    \includegraphics[scale = 0.38]{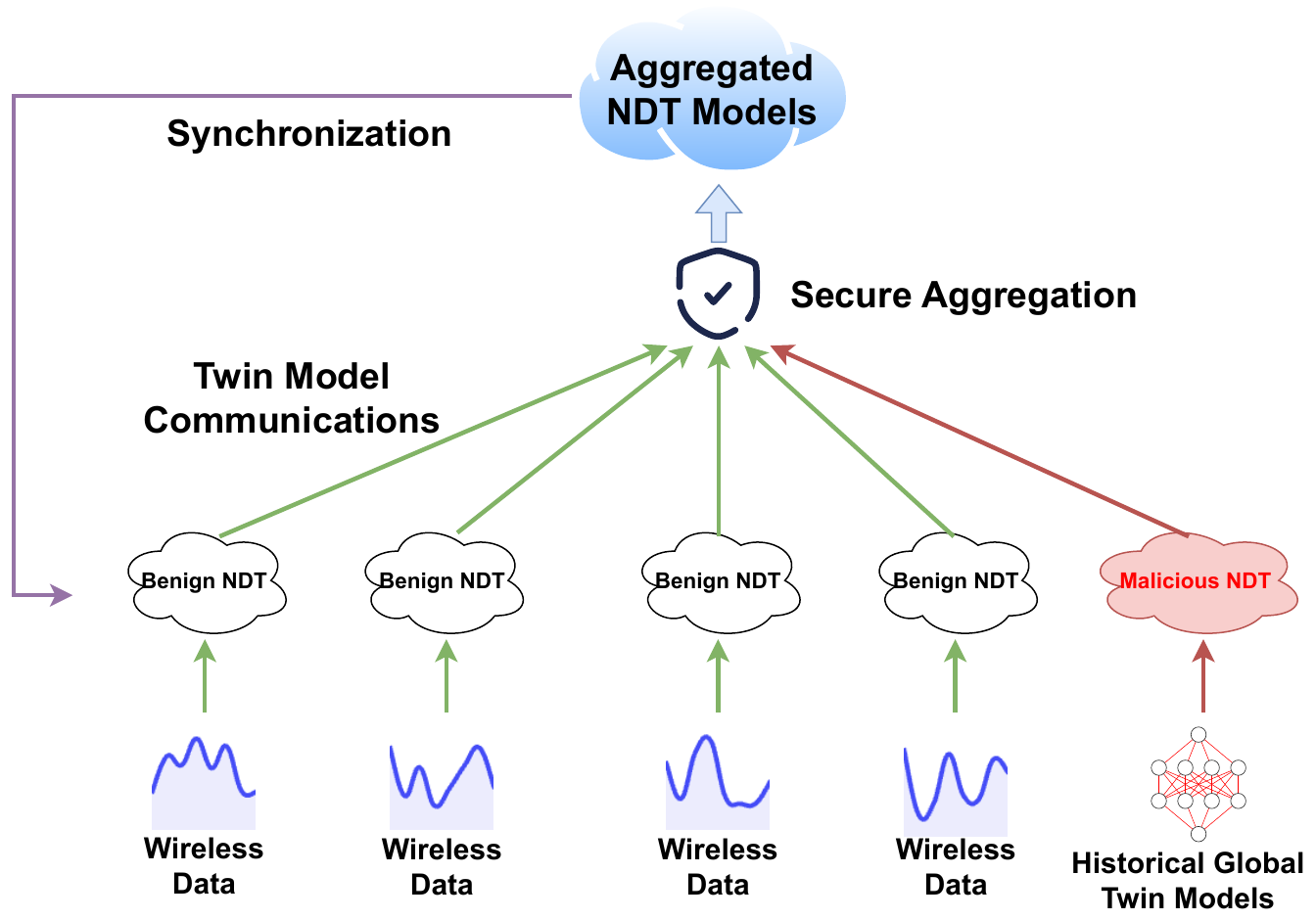}
    \caption{Framework of NDT system protections.}
    \label{fig:security}
\end{figure}

To achieve their goal, attackers introduce counterfeit NDT models into the system, as illustrated in Fig.~\ref{fig:security}. These fabricated NDTs can replicate the functionality of legitimate NDTs with minimal investment and effort. This tactic, which entails deploying fake BSs and NDTs using readily available open-source tools or emulators~\cite{simulatefake,NoxPlayer,bluestacks,cao2022mpaf}, presents a low-barrier, high-feasibility threat vector distinct from the strategies like those in~\cite{zheng2022poisoning} that require compromising actual NDTs. Given the stringent security measures of contemporary networks, which complicate the direct manipulation of authentic twin models, this approach of deploying spurious BSs and NDTs emerges as a notably viable method for attack.
\vspace{-0.15in}
\subsection{Knowledge of the Attacker}
The attacker's limited understanding of the intricacies of the targeted distributed NDT system adds to the challenge of mounting a successful attack. In many practical scenarios, acquiring comprehensive knowledge about the aggregation algorithms or details of legitimate NDTs proves exceedingly difficult due to robust security measures and encryption. Consequently, an attack necessitating minimal specialized knowledge and training data not only appears more feasible but also carries a lower risk of detection. The operation of the counterfeit NDTs—receiving the global model and dispatching malicious updates—demands only basic intelligence, effectively lowering the threshold for entry for would-be attackers. This characteristic of the threat model heightens its potential danger, broadening the pool of possible adversaries to include those with scant technical skills or resources.
\vspace{-0.15in}
\subsection{Fake Traffic Injection Attack}

The proposed Algorithm~\ref{algo:attack}, named the Fake Traffic Injection (FTI) Algorithm, presents a strategy for a Byzantine model poisoning attack aimed at compromising the prediction accuracy of an NDT system under specific assumptions.

At the core of the FTI attack is an iterative procedure. In each iteration, the current global twin model $\bm{\theta}^{t}$ and the base model $\hat{\bm{\theta}}$ undergo a detailed examination. For each fake Base Station (BS) $i$, a malicious local model $\bm{\theta}_i^{t}$ is constructed by blending the global model $\bm{\theta}^{t}$ with the base model $\hat{\bm{\theta}}$ in a weighted manner, as delineated in line 5 of Algorithm~\ref{algo:attack}. Subsequent to the formation of $\bm{\theta}_i^{t}$, its deviation from the global model is assessed using the Euclidean norm, as depicted in line 7. The algorithm then evaluates whether this distance has increased compared to the previous measurement, denoted as \textit{PreDist}. If an increase is observed, indicating that the malicious local model $\bm{\theta}_i^{t}$ is diverging further from the global model $\bm{\theta}^{t}$, the value of $\eta$ is incremented. Conversely, if no increase in distance is detected, $\eta$ is decremented. The adjustment of $\eta$ is executed in half-steps of its initial value, as outlined in lines 8 to 12.

The algorithm aims to steer the global model towards greater alignment with a pre-defined base model in each round. Specifically, during the $t$-th round, fake NDTs compute the direction of local model updates, determined by the difference between the current global twin model and the base model, denoted as $\bm{H} = \hat{\bm{\theta}} - \bm{\theta}^t$. Progressing in this direction signifies that the global model is becoming more akin to the base model. A straightforward method to obtain the local model of a fake BS involves scaling $\bm{H}$ by a factor $\eta$. However, this direct approach yields suboptimal attack performance.

\begin{figure}
    \centering
    \includegraphics[scale = 0.30]{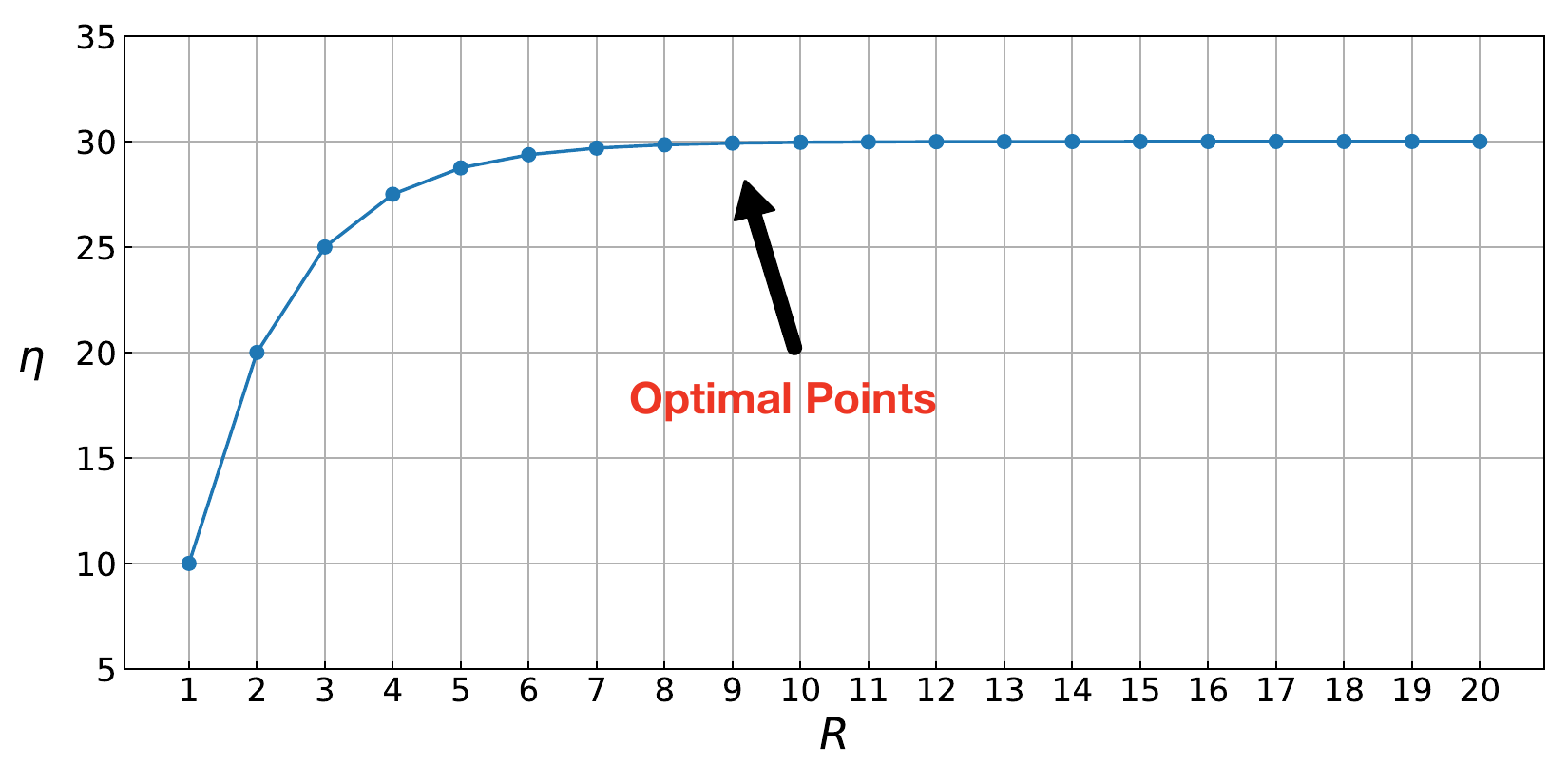}
    \caption{Optimal value of $\eta$ over communication round of $R$.}
    \label{fig:eta_r}
\end{figure}

Assuming $n$ represents the number of benign NDTs, and the attacker intends to inject $m$ fake NDTs into the system, we propose a method for calculating $\bm{\theta}_i^{t}$ for each fake NDT $i \in [n+1, n+m]$:
\begin{equation}
\bm{\theta}^t_i = \eta \hat{\bm{\theta}} - (\eta - 1)\bm{\theta}^{t}.
\end{equation}
In such scenarios, an attacker tends to opt for a higher $\eta$ to ensure the sustained effectiveness of the attack, as illustrated in Fig.~\ref{fig:eta_r} with an initial $\eta$ of 10. This remains valid even after the server amalgamates the manipulated local updates from fake NDTs with legitimate updates from benign NDTs.

\begin{algorithm}[t]
\caption{Fake Traffic Injection (FTI)}
\begin{algorithmic}[1]
\Require Current global twin model $\bm{\theta}^{t}$, base model $\hat{\bm{\theta}}$, $n$ benign NDTs, $m$ fake NDTs, $\eta$
\Ensure Fake models $\bm{\theta}_i^{t}, i \in [n+1, n+m]$
\State $\text{step} \gets \eta$
\State $\text{PreDist} \gets -1$
\For{$r = 1, 2, \ldots, R$}
    \For{each fake NDT $i$}
        \State $\bm{\theta}_i^{t} \gets \eta \hat{\bm{\theta}} - (\eta - 1)\bm{\theta}^{t}$
    \EndFor
    \State $\text{Dist} \gets \left\| \bm{\theta}_i^{t} - \bm{\theta}^t \right\|_2$
    \If{$\text{PreDist} < \text{Dist}$}
        \State $\eta \gets \eta + \frac{\text{step}}{2}$
    \Else
        \State $\eta \gets \eta - \frac{\text{step}}{2}$
    \EndIf
    \State $\text{step} \gets \frac{\text{step}}{2}$
    \State $\text{PreDist} \gets \text{Dist}$
\EndFor
\State \Return $\bm{\theta}_i^{t}, i \in [n+1, n+m]$
\end{algorithmic}
\label{algo:attack}
\end{algorithm}

\section{Global-Local Inconsistency Detection}
\label{sec:defense}

The defense against model poisoning attacks is founded on an aggregation protocol designed to identify malicious NDTs, termed the Global-Local Inconsistency Detection (GLID) method, as elaborated in Algorithm~\ref{algo:defense}. In each global round $t$, GLID primarily examines anomalies present in each dimension of the model parameters $\bm{\theta}_i^t$, aiding in the identification of potentially malicious entities, where $i \in [1, n+m]$ and $n+m$ denotes the total number of NDTs in the system. This robust and versatile approach enables the system to adapt to various operational contexts without necessitating intricate similarity assessments.
Then, choosing the parameter for the percentile range when trimming outliers for secure aggregation becomes crucial, as it directly influences the model's balance between robustness and accuracy. Typically, a narrow percentile range might exclude legitimate variations in data, reducing the model's accuracy and potentially leading to biased or incomplete representations. Conversely, a broad percentile range may fail to eliminate malicious or anomalous data contributions, compromising the model's security by allowing adversarial inputs to skew the aggregation process. Therefore, selecting an appropriate percentile range ensures that most benign data points are retained while effectively filtering out outliers or adversarial inputs. This balance is essential for maintaining both the performance and security of digital twin models, protecting against data poisoning attacks without sacrificing the overall quality and representativeness of the aggregated twinning data.

Specifically, the GLID approach enhances the detection of potential malicious activities within the network by employing \textit{percentile-based trimming} on each dimension of the model parameters. To establish an effective percentile pair for identifying abnormalities, four statistical methods can be adopted: Standard Deviation (SD), Interquartile Range (IQR), z-scores, or One-class Support Vector Machine (One-class SVM).

Suppose the total count of dimensions of the model parameter is $D$. For the default SD method, the percentile pair for each dimension $d$ can be calculated as follows:
\begin{equation}
    \text{percentile pair}^t_d = \left( g\left( \bar{\bm{\theta}}_d^t - k \cdot \sigma_d^t\right), \, g\left( \bar{\bm{\theta}}_d^t + k \cdot \sigma_d^t \right) \right),
\end{equation}
where $\bar{\bm{\theta}}_d^t$ is the mean of the $d$-th dimension across all models in the $t$-th global training round, $\sigma_d^t$ is the standard deviation of the $d$-th dimension, and $k$ is a predefined constant dictating the sensitivity of outlier detection. $g(\cdot)$ is the interpolation function based on the standard deviation bound to estimate percentile pairs, defined as:
\begin{equation}
    g(x) = \left( \frac{P(x) - 0.5}{n+m} \right) \times 100,
\end{equation}
where $P(x)$ is the position of $x$ in the sorted dataset. We use $k=3$ for general purposes. Given that different tasks may require varied percentile bounds, a precise estimation method is crucial for generalizing our defense strategy. The detailed percentile estimation methods are discussed later in this section. In the FL-based WTP system, model parameters in the $d$-th dimension exceeding these percentile limits are flagged as malicious, and their weights $\alpha^t_i$ are assigned as 0. The other benign values in this dimension are aggregated using a weighted average rule, where the weights $\alpha^t_{d,i}$ are inversely proportional to the absolute deviation of each value $\bm{\theta}^t_{d,i}$ from the mean $\bar{\bm{\theta}}_d^t$, and normalized by the standard deviation $\sigma_d^t$. It can be represented as follows:
\begin{equation}
    \alpha^t_{d,i} = \frac{\sigma_d^t}{\left| \bm{\theta}^t_{d,i} - \bar{\bm{\theta}}_d^t \right|}.
\end{equation}
These weights of the $d$-th dimension are then normalized and applied to aggregate each BS's local model $\bm{\theta}_i^t$ into a global model $\bm{\theta}^{t+1}$, which can be represented as follows in the view of each dimension:
\begin{equation}
    \bm{\theta}^{t+1}_d = \frac{\sum_{i=1}^{n+m} \alpha^t_{d,i} \cdot \bm{\theta}^t_{d,i}}{\sum_{i=1}^{n+m} \alpha^t_{d,i}}.
\end{equation}
Subsequently, the server broadcasts this aggregated global model parameter $\bm{\theta}^{t+1}$ back to all NDTs for synchronization.

There are three additional percentile estimation strategies listed below. Based on the upper and lower bound computed below, we can get a final percentile estimation decision to detect abnormal values in each dimension.

\begin{itemize}
    \item \textbf{Interquartile Range (IQR)}: 
    The IQR method calculates the range between the first and third quartiles (25th and 75th percentiles) of the data, identifying outliers based on this range. For each dimension $d$, the outlier bounds are:
    \begin{align}
        \text{lower bound}^t_{d,\text{IQR}} &= Q1^t_d - k_{\text{IQR}} \cdot \text{IQR}^t_d, \\
        \text{upper bound}^t_{d,\text{IQR}} &= Q3^t_d + k_{\text{IQR}} \cdot \text{IQR}^t_d,
    \end{align}
    where $Q1^t_d$ and $Q3^t_d$ are the first and third quartiles, and $k_{\text{IQR}}$ adjusts sensitivity.

    \item \textbf{Z-scores}: 
    The Z-score method measures how many standard deviations a point is from the mean. For each dimension $d$, the normal range bounds are:
    \begin{align}
        \text{lower bound}^t_{d,\text{Z-score}} &= g\left( \bar{\bm{\theta}}_d^t - k_{\text{Z}} \cdot \sigma_d^t\right), \\
        \text{upper bound}^t_{d,\text{Z-score}} &= g\left( \bar{\bm{\theta}}_d^t + k_{\text{Z}} \cdot \sigma_d^t \right),
    \end{align}
    where $k_{\text{Z}}$ is the number of standard deviations for the normal range.

    \item \textbf{One-Class SVM}: 
    One-class SVM constructs a decision boundary for anomaly detection. The decision function for each dimension $d$ is:
    \begin{align}
        f^t_d(\bm{\theta}) &= \text{sign}\left( \sum_{i=1}^{n_{\text{SV}}} \gamma_i \cdot K(\bm{\theta}^t_{\text{SV}_i, d}, \bm{\theta}) - \rho \right), \\
        \text{where} \, & \bm{\theta}^t_{\text{SV}_i, d} \, \text{are the support vectors,} \nonumber \\
        & \gamma_i \, \text{are the Lagrange multipliers,} \nonumber \\
        & K(\cdot, \cdot) \, \text{is the kernel function, and} \nonumber \\
        & \rho \, \text{is the offset.} \nonumber
    \end{align}
    A point $\bm{\theta}$ is an outlier if $f^t_d(\bm{\theta}) < 0$.
\end{itemize}

In essence, this defense mechanism is a strategic amalgamation of direct statistical trimming and aggregation, targeting the preservation of the global model's integrity against poisoning attacks. By accurately isolating and excluding malicious NDTs prior to aggregation, it significantly diminishes the likelihood of adversarial disruption in the FL framework. Additionally, its capacity to accommodate various dimensions and adapt to different inconsistency metrics and aggregation protocols considerably extends its applicability across a broad spectrum of distributed wireless network scenarios.

\begin{algorithm}[t]
\caption{Global-local Inconsistency Detection (GLID)}
\begin{algorithmic}[1]
\Require Local models \( \bm{\theta}_1^t, \bm{\theta}_2^t, \ldots, \bm{\theta}_{n+m}^t \), current global model \( \bm{\theta}^t \), \(k\)
\Ensure Aggregated global model \( \bm{\theta}^{t+1} \)

\For{\( d = 1, 2, \ldots, D \)}
    \State \( \bar{\bm{\theta}}_d^t \gets \frac{1}{n+m} \sum_{i=1}^{n+m} \bm{\theta}^t_{d,i} \)  
    \State \( \sigma_d^t \gets \sqrt{\frac{1}{n+m} \sum_{i=1}^{n+m} (\bm{\theta}^t_{d,i} - \bar{\bm{\theta}}_d^t)^2} \)  
    \State \( \text{percentile}^t_d \gets \left( g\left( \bar{\bm{\theta}}_d^t - k \cdot \sigma_d^t\right), \, g\left( \bar{\bm{\theta}}_d^t + k \cdot \sigma_d^t \right) \right) \)
    \State Identify malicious NDTs based on percentile pairs
    \For{each NDT $i$}
        \If{\( \bm{\theta}^t_{d,i} \) is benign}
            \State \( \alpha^t_{d,i} \gets \frac{\sigma_d^t}{\left| \bm{\theta}^t_{d,i} - \bar{\bm{\theta}}_d^t \right|} \)
        \Else
            \State \( \alpha^t_{d,i} \gets 0 \)
        \EndIf
    \EndFor
    \State \( \bm{\theta}_d^{t+1} \gets \frac{\sum_{i=1}^{n+m} \alpha^t_{d,i} \cdot \bm{\theta}^t_{d,i}}{\sum_{i=1}^{n+m} \alpha^t_{d,i}} \)
\EndFor
\State \(\bm{\theta}^{t+1} \gets \left[ \bm{\theta}_1^{t+1}, \bm{\theta}_2^{t+1}, \ldots, \bm{\theta}_D^{t+1} \right]\)
\State \Return \( \bm{\theta}^{t+1} \)
\end{algorithmic}
\label{algo:defense}
\end{algorithm}
\section{Experimental Evaluation}
\label{sec:evaluation}

In this section, we present an extensive evaluation of our proposed FTI poisoning attack and the GLID defense mechanism. We provide extensive results across various performance metrics to demonstrate their effectiveness in multiple dimensions.

\begin{table*}[ht]
\centering
\caption{Performance Evaluation with Milan-Internet Dataset during V-twinning Stage}
\label{tab:net-milan_v}
\begin{tabular}{|c|c|C{1.2cm}|C{1.2cm}|C{1.2cm}|C{1.2cm}|C{1.2cm}|C{1.2cm}|C{1.2cm}|}
\hline
\multirow{2}{*}{Aggregation Rule} & \multirow{2}{*}{Metric} & \multicolumn{7}{c|}{Attack} \\ \cline{3-9}
                                  &                        & NO & Trim & History & Random & MPAF & Zheng & \textit{\textbf{FTI}} \\ \hline
\multirow{2}{*}{Mean}
& MAE & 0.281 & 100.0 & 100.0 & 100.0 & 100.0 & 0.768 & \textbf{100.0} \\
& MSE & 0.106 & 100.0 & 100.0 & 100.0 & 100.0 & 0.314 & \textbf{100.0} \\ \hline
\multirow{2}{*}{Median}
& MAE & 0.281 & 0.283 & 0.281 & 0.282 & 0.281 & 0.287 & \textbf{100.0} \\
& MSE & 0.106 & 0.106 & 0.107 & 0.106 & 0.106 & 0.115 & \textbf{100.0} \\ \hline
\multirow{2}{*}{Trim}
& MAE & 0.281 & 0.282 & 0.282 & 0.281 & 0.282 & 0.309 & \textbf{100.0} \\
& MSE & 0.106 & 0.107 & 0.109 & 0.106 & 0.108 & 0.126 & \textbf{100.0} \\ \hline
\multirow{2}{*}{Krum}
& MAE & 0.291 & 0.295 & 100.0 & 0.295 & 100.0 & 0.295 & \textbf{100.0} \\
& MSE & 0.111 & 0.113 & 100.0 & 0.114 & 100.0 & 0.114 & \textbf{100.0} \\ \hline
\multirow{2}{*}{FoolsGold}
& MAE & 0.283 & 100.0 & 100.0 & 100.0 & 100.0 & 1.004 & \textbf{100.0} \\
& MSE & 0.115 & 100.0 & 100.0 & 100.0 & 100.0 & 0.627 & \textbf{100.0} \\ \hline
\multirow{2}{*}{FABA}
& MAE & 0.289 & 100.0 & 0.297 & 0.289 & 100.0 & 0.693 & \textbf{100.0} \\
& MSE & 0.109 & 100.0 & 0.105 & 0.101 & 100.0 & 0.269 & \textbf{100.0} \\ \hline
\multirow{2}{*}{FLTrust}
& MAE & 0.312 & 0.304 & 100.0 & 0.310 & 100.0 & 3.252 & \textbf{100.0} \\
& MSE & 0.114 & 0.112 & 100.0 & 0.114 & 100.0 & 1.278 & \textbf{100.0} \\ \hline
\multirow{2}{*}{FLAIR}
& MAE & 0.286 & 0.298 & 100.0 & 100.0 & 100.0 & 0.320 & \textbf{100.0} \\
& MSE & 0.114 & 0.108 & 100.0 & 100.0 & 100.0 & 0.116 & \textbf{100.0} \\ \hline
\multirow{2}{*}{\textbf{GLID}}
& MAE & \textbf{0.281} & \textbf{0.281} & \textbf{0.282} & \textbf{0.281} & \textbf{0.281} & \textbf{0.282} & \textbf{72.453} \\
& MSE & \textbf{0.106} & \textbf{0.107} & \textbf{0.106} & \textbf{0.106} & \textbf{0.107} & \textbf{0.106} & \textbf{27.548} \\ \hline

\end{tabular}
\end{table*}

\subsection{Experimental Setup}

\subsubsection{Datasets}
To assess our methods, we employ real-world datasets from Telecom Italia~\cite{dataset}. The Milan wireless traffic dataset is partitioned into 10,000 grid cells, each served by an NDT covering an area of approximately 235 meters squared. The dataset comprises three subsets: ``Milan-Internet'', ``Milan-SMS'', and ``Milan-Calls'', which capture diverse wireless usage patterns. Our primary focus is on the ``Milan-Internet'' subset, which facilitates a detailed analysis of urban telecommunications behavior.

\subsubsection{Baseline Schemes}
We benchmark our FTI attack against several state-of-the-art model poisoning attacks to underscore its effectiveness. Additionally, we employ these baseline attacks to demonstrate the efficacy of our GLID defense strategy:

\begin{itemize}
    \item \textbf{Trim attack~\cite{fang2020local}:} Processes each key in a model dictionary, using extremes in a specific dimension to determine a \textit{directed} dimension. Model parameters are then selectively zeroed or retained to influence the model's behavior.
    
    \item \textbf{History attack~\cite{cao2022mpaf}:} Iterates over model parameters, replacing current values with historically scaled ones to warp the model parameters using past data and misguide the aggregation process.
    
    \item \textbf{Random attack~\cite{cao2022mpaf}:} Disrupts the model by replacing parameters with random values drawn from a normal distribution, scaled to maintain a semblance of legitimacy and inject controlled chaos into the aggregation process.
    
    \item \textbf{MPAF~\cite{cao2022mpaf}:} Calculates a directional vector from the difference between initial and current parameters, adjusting model values to intentionally diverge from the original trajectory and introduce adversarial bias. Fake NDTs are then injected into the system.
    
    \item \textbf{Zheng attack~\cite{zheng2022poisoning}:} Inverts the direction of model updates by incorporating the negative of previous global updates, refined through error maximization to generate a poison that is challenging to detect due to its alignment with the twin model's error landscape.
\end{itemize}

Furthermore, we consider several baseline defensive mechanisms to evaluate the robustness of our proposed attack and defense:

\begin{itemize}
    \item \textbf{Mean~\cite{mcmahan2017communication}:} Calculates the arithmetic mean of updates in each dimension, assuming equal trustworthiness among all NDTs. This method is susceptible to the influence of extreme values.
    
    \item \textbf{Median~\cite{yin2018byzantine}:} Identifies the median value in each dimension for each parameter across updates, discarding extreme contributions to enhance robustness against outliers.
    
    \item \textbf{Trim~\cite{yin2018byzantine}:} Discards a specified percentage of the highest and lowest updates before computing the mean in each dimension, reducing the influence of anomalous or malicious updates on the aggregate model.
    
    \item \textbf{Krum~\cite{blanchard2017machine}:} Scores each NDT's update based on the sum of Euclidean distances to other NDTs' updates, selecting the update from the NDT with the minimum score for the global update.
    
    \item \textbf{FoolsGold~\cite{fung2018mitigating}:} Calculates a cosine similarity matrix among all NDTs and adjusts the weights for each NDT based on these similarities, aggregating the weighted gradients to form a global model.
    
    \item \textbf{FABA~\cite{xia2019faba}:} Computes the Euclidean distance for each NDT's model from the mean of all received models, excluding a specific percentage of the most distant models to filter out potential outliers or malicious updates.
    
    \item \textbf{FLTrust~\cite{cao2020fltrust}:} Calculates cosine similarity between the server's current model and each NDT's model to generate trust scores, which are then used to weigh the NDT's contribution to the final aggregated model.
    
    \item \textbf{FLAIR~\cite{sharma2023flair}:} Each NDT calculates ``flip-scores'' from the changes in gradient directions and ``suspicion-scores'' based on historical behavior, using these scores to adjust the weights assigned to each NDT's contributions to the global twin model.
\end{itemize}

\subsubsection{Experimental Settings and Performance Metrics}

For our experiments, we randomly select 100 BSs and their corresponding NDTs to evaluate the impact of poisoning attacks and the effectiveness of defense mechanisms. We primarily report results on the Milan-Internet dataset. Model training is configured with a learning rate of 0.001 and a batch size of 64. We inject a 20\% percentage of fake NDTs to simulate benign ones in the system for the FTI attack and assume a scenario where 20\% of the NDTs are compromised for other baseline attacks. Our proposed FTI attack employs a parameter \( \eta = 10 \), while other attacks utilize a scaling factor of 1000. For the Trim aggregation rule, we discard 20\% of the twin model parameters from all NDTs. In our GLID defense, we use the standard deviation (SD) method as the default percentile estimation method. We adopt Mean Absolute Error (MAE) and Mean Squared Error (MSE) as the primary metrics for performance evaluation, with larger MAE and MSE values indicating better attack effectiveness.

\subsection{Numerical Results}

\subsubsection{Performance of Proposed Methods}

\begin{table*}[ht]
\centering
\caption{Performance Evaluation with Milan-Internet Dataset during H-twinning Stage}
\label{tab:net-milan_h}
\begin{tabular}{|c|c|C{1.2cm}|C{1.2cm}|C{1.2cm}|C{1.2cm}|C{1.2cm}|C{1.2cm}|C{1.2cm}|}
\hline
\multirow{2}{*}{Aggregation Rule} & \multirow{2}{*}{Metric} & \multicolumn{7}{c|}{Attack} \\ \cline{3-9}
& & NO & Trim & History & Random & MPAF & Zheng & \textit{\textbf{FTI}} \\ \hline
\multirow{2}{*}{Mean}
& MAE & 0.266 & 100.0 & 100.0 & 100.0 & 100.0 & 0.753 & \textbf{100.0} \\
& MSE & 0.101 & 100.0 & 100.0 & 100.0 & 100.0 & 0.309 & \textbf{100.0} \\ \hline
\multirow{2}{*}{Median}
& MAE & 0.296 & 0.298 & 0.296 & 0.297 & 0.296 & 0.302 & \textbf{100.0} \\
& MSE & 0.101 & 0.102 & 0.102 & 0.105 & 0.101 & 0.110 & \textbf{100.0} \\ \hline
\multirow{2}{*}{Trim}
& MAE & 0.296 & 0.297 & 0.297 & 0.296 & 0.297 & 0.294 & \textbf{100.0} \\
& MSE & 0.101 & 0.102 & 0.104 & 0.101 & 0.108 & 0.121 & \textbf{100.0} \\ \hline
\multirow{2}{*}{Krum}
& MAE & 0.276 & 0.280 & 100.0 & 0.280 & 100.0 & 0.280 & \textbf{100.0} \\
& MSE & 0.106 & 0.108 & 100.0 & 0.109 & 100.0 & 0.109 & \textbf{100.0} \\ \hline
\multirow{2}{*}{FoolsGold}
& MAE & 0.268 & 100.0 & 100.0 & 100.0 & 100.0 & 0.989 & \textbf{100.0} \\
& MSE & 0.110 & 100.0 & 100.0 & 100.0 & 100.0 & 0.622 & \textbf{100.0} \\ \hline
\multirow{2}{*}{FABA}
& MAE & 0.274 & 100.0 & 100.0 & 100.0 & 100.0 & 0.678 & \textbf{100.0} \\
& MSE & 0.104 & 100.0 & 100.0 & 100.0 & 100.0 & 0.264 & \textbf{100.0} \\ \hline
\multirow{2}{*}{FLTrust}
& MAE & 0.297 & 0.289 & 100.0 & 0.295 & 100.0 & 3.237 & \textbf{100.0} \\
& MSE & 0.109 & 0.107 & 100.0 & 0.109 & 100.0 & 1.223 & \textbf{100.0} \\ \hline
\multirow{2}{*}{FLAIR}
& MAE & 0.271 & 0.283 & 100.0 & 100.0 & 100.0 & 0.305 & \textbf{100.0} \\
& MSE & 0.109 & 0.103 & 100.0 & 100.0 & 100.0 & 0.111 & \textbf{100.0} \\ \hline
\multirow{2}{*}{\textbf{GLID}}
& MAE & \textbf{0.266} & \textbf{0.266} & \textbf{0.267} & \textbf{0.266} & \textbf{0.266} & \textbf{0.267} & \textbf{72.458} \\
& MSE & \textbf{0.101} & \textbf{0.102} & \textbf{0.101} & \textbf{0.101} & \textbf{0.102} & \textbf{0.101} & \textbf{27.543} \\ \hline

\end{tabular}
\end{table*}

Tables I and II demonstrate the significant vulnerabilities introduced by the proposed FTI Attack across various aggregation methods within our NDT construction.
It is observed that under our FTI Attack, the Mean Rule is completely compromised over both the V-twinning and H-twinning stages, as reflected by their MAE and MSE values reaching over 100.0 (values exceeding 100 are capped at 100). This result denotes a total breakdown in their wireless traffic prediction functionality. The Median Rule further emphasizes the severity of the FTI Attack, with both its MAE and MSE escalating from modest baseline figures to 100. This sharp contrast highlights the FTI attack's reliable performance against other defenses, such as the Trim Attack against the Median Rule, where the increase in MAE and MSE is relatively minor at 0.283 and 0.106 for V-twinning, respectively.
Additionally, the Trim Rule, typically considered robust, exhibits a drastic increase in MAE to over 100.0, a significant rise from its baseline without any attack (denoted as NO in Tables \ref{tab:net-milan_v} and \ref{tab:net-milan_h}) of 0.281. This surge underscores the Trim Rule's vulnerability to the FTI Attack, marking a notable departure from its typical resilience. Similar results can also be found in other aggregation rules under FTI attacks, such as Krum, FoolsGold, FABA, FLTrust, and FLAIR, where the FTI attack demonstrates the best overall performance against the given defenses.
The Zheng Attack, however, presents a distinct pattern of disruption. When subjected to this attack, FLTrust, which typically exhibits lower error metrics, shows a significant compromise, evidenced by the dramatic increase in its MAE to 3.252 and MSE to 1.278. Such a tailored nature of the Zheng Attack appears to target specific vulnerabilities within FLTrust, which are not as apparent in other scenarios, such as the Trim Attack, where the rise in MAE and MSE for FLTrust is relatively modest. Regarding the MPAF Attack, most aggregation rules in the table do not show a convincing defense, except for a few like Median, Trim, and GLID.

During the H-twinning stage, most baseline schemes demonstrate a similar performance under attacks. However, although the Median and Trim Rules could protect the NDT systems from being attacked, they do not perform well in maintaining a precise NDT after a valid initial construction, i.e. the V-twinning stage. For instance, the MAE and MSE values of Median Rule under NO attacks are 0.281 and 0.106, respectively. These performance metrics increase to 0.296 and 0.101 after a period of maintenance, which leads to inaccurate predictions compared to the initial twin models. This is due to the heterogeneous nature of data distribution, with the Median and Trim Rules trimming out too many participants during the twin model aggregation process.

From the defender's standpoint, the proposed GLID aggregation method demonstrates consistent performance stability across various attacks. Both its MAE and MSE values remain close to their baseline levels. Even in the case of our FTI attack, GLID manages to keep errors below 100, with MAE and MSE values of 72.453 and 27.548, respectively. This stability is particularly noteworthy, especially when compared to other rules such as FLAIR, which exhibit a significant deviation from their non-attacked baselines under the same adversarial conditions. GLID's ability to sustain its performance in the face of diverse and severe attacks underscores its potential as a resilient aggregation methodology.
Other rules, such as FABA, also experience inconsistent defense performance during the V-twinning and H-twinning stages. FABA maintains good performance under History and Random attacks during the V-twinning process but is degraded to over 100.0 during the H-twinning process. Such performance is led by the various data distribution and sample sizes from the real-time data stream. In later evaluations, we focus on the entire twinning process, combining V-twinning and H-twinning, to evaluate the effects of other parameters on our proposed poisoning attacks and defense mechanisms for NDT systems.

\subsubsection{Evaluation on the Impact of $\eta$}

\begin{figure}[ht]
    \centering
    \includegraphics[scale = 0.35]{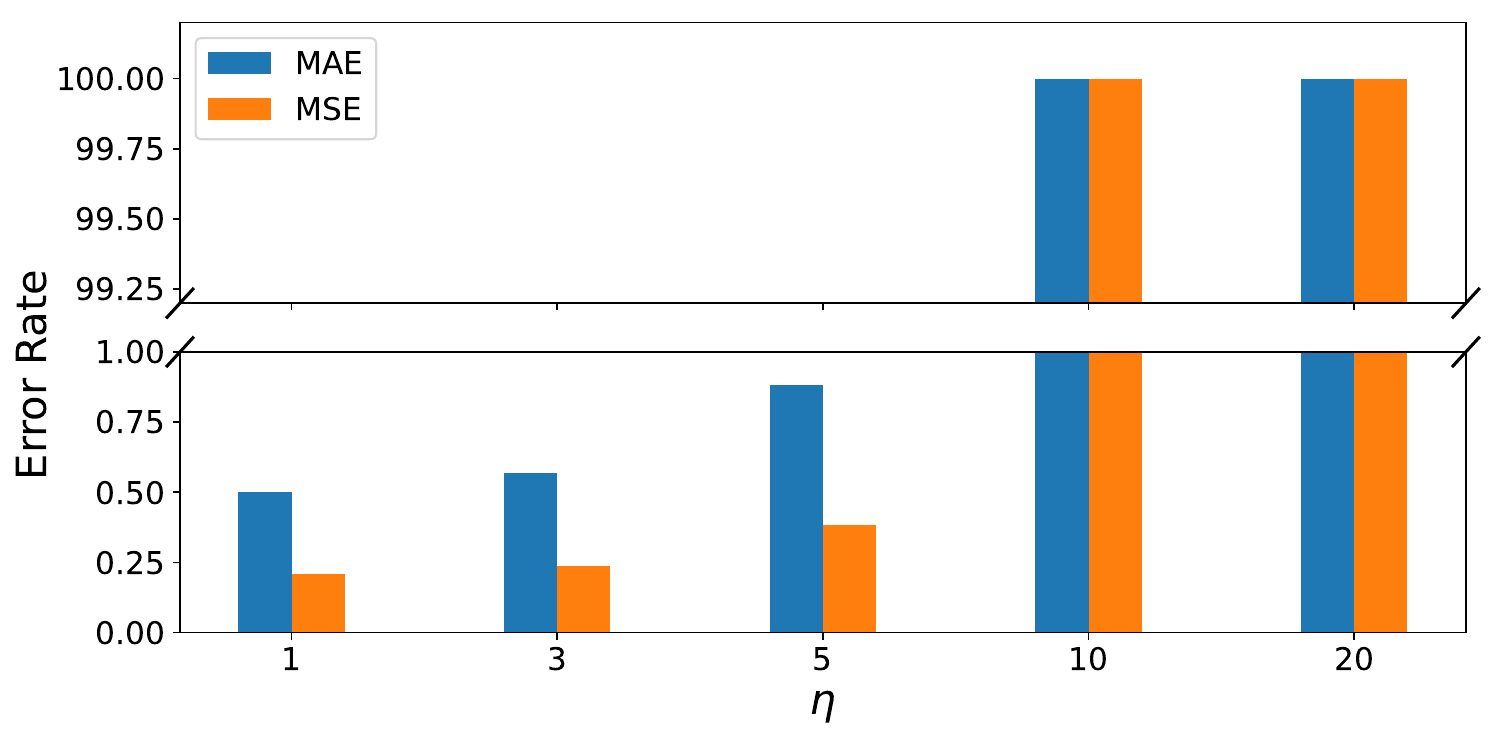}
    \caption{Impact of Values of \( \eta \).}
    \label{fig:eta}
    \vspace{-0.2cm}
\end{figure}

The step size \( \eta \) in our proposed FTI attack (see Algorithm~\ref{algo:attack}) serves as a dynamic scaling factor, and its initial value significantly influences the NDT's performance metrics. This impact is illustrated in Fig.~\ref{fig:eta}, where the Median aggregation rule is employed as the baseline defense strategy. A notable observation is the correlation between increasing values of \( \eta \) and the corresponding rise in MAE and MSE of twin models. For example, at \( \eta = 1 \), the MAE and MSE are relatively low, recorded at 0.517 and 0.215, respectively. However, increasing \( \eta \) to higher values, such as 10 or 20, results in a dramatic surge that reaches the maximum error rate. This increase suggests a significant compromise in the twin models, surpassing the predefined threshold for effective detection of the attack.
The rationale behind this analysis emphasizes the pivotal role of \( \eta \) in determining the \textit{strength} of a poisoning attack. An increased initial \( \eta \) tends to degrade model performance, deviating significantly from its expected operational state. Simultaneously, a higher \( \eta \) also raises the risk of the attack's perturbations being detected and eliminated during the defense process.

\subsubsection{Evaluation on Percentage of Fake NDTs}

The degree of compromise in NDTs significantly influences the model's performance, as evidenced in Table~\ref{tab:percentage}. By adopting the Median aggregation as the defensive approach, the model first exhibits resilience at lower compromise levels, such as with only 5\%--10\% fake NDTs in the scenario. However, a noticeable decline in performance is observed as the percentage of fake NDTs increases to 20\% or higher. This deterioration is evident as the MAE and MSE values reach 100.0 in all categories, signaling a complete model failure.
The underlying principle behind this trend suggests the model's limited tolerance to malicious interference. More precisely, the network system can withstand below 20\% compromise without significant performance degradation. However, beyond this threshold, the model's integrity is severely undermined, resulting in a complete system breakdown. This observation highlights the critical importance of implementing robust security measures to prevent excessive compromise of NDTs, ensuring the model's reliability and effectiveness.

\begin{table}[ht]
\centering
\caption{Impact of Percentages of Fake NDTs}
\label{tab:percentage}
\scriptsize
\begin{tabular}{|c|c|c|c|c|c|c|c|}
\hline
\multirow{2}{*}{Pct.} & \multirow{2}{*}{Metric} & \multicolumn{6}{c|}{Attack} \\
\cline{3-8}
& & {Trim} & {Hist} & {Rand} & {MPAF} & {Zhe.} & {\textbf{FTI}} \\
\hline
\multirow{2}{*}{5\%} & MAE & 0.276 & 0.270 & 0.274 & 0.270 & 0.268 & \textbf{0.284} \\
& MSE & 0.093 & 0.094 & 0.093 & 0.093 & 0.093 & \textbf{0.094} \\
\hline
\multirow{2}{*}{10\%} & MAE & 0.275 & 0.268 & 0.273 & 0.268 & 0.269 & \textbf{0.313} \\
& MSE & 0.092 & 0.095 & 0.093 & 0.095 & 0.101 & \textbf{0.109} \\
\hline
\multirow{2}{*}{20\%} & MAE & 0.278 & 0.273 & 0.273 & 0.271 & 0.324 & \textbf{100.0} \\
& MSE & 0.092 & 0.101 & 0.092 & 0.097 & 0.141 & \textbf{100.0} \\
\hline
\multirow{2}{*}{30\%} & MAE & 100.0 & 100.0 & 100.0 & 100.0 & 6.045 & \textbf{100.0} \\
& MSE & 100.0 & 100.0 & 6.146 & 100.0 & 1.159 & \textbf{100.0} \\
\hline
\multirow{2}{*}{40\%} & MAE & 100.0 & 100.0 & 100.0 & 100.0 & 100.0 & \textbf{100.0} \\
& MSE & 100.0 & 100.0 & 100.0 & 100.0 & 100.0 & \textbf{100.0} \\
\hline
\end{tabular}
\end{table}

\subsubsection{Evaluations on Percentile Estimation Methods}

The dynamic trimming of an adaptive number of model parameters through percentile estimation, which is adapted in GLID, proves to be an effective defense strategy against various model poisoning attacks. In the comparative analysis of various estimation methods, as shown in Table~\ref{tab:estimation-method}, SD estimation emerges as the best technique, exhibiting marked consistency and robustness across a spectrum of estimation approaches. This is evidenced by the consistently low MAE and MSE values for SD across these approaches, at 0.219 and 0.087, respectively.
In contrast, other methods have varying degrees of inconsistency and vulnerability. For instance, One-class SVM exhibits pronounced variability, with MAE and MSE values reaching the maximal error level of over 100.0 under Trim, History, and MPAF attacks. Such a disparity in performance, particularly the stably lower error rates of SD compared to the significant fluctuations in other estimation methods, positions SD as a reliable and effective percentile estimation technique in GLID.

\begin{table}
\centering
\caption{Impact of Percentile Estimation Methods}
\label{tab:estimation-method}
\setlength{\tabcolsep}{4pt} 
\scriptsize
\begin{tabular}{|c|c|c|c|c|c|c|c|c|}
\hline
\multirow{2}{*}{Method} & \multirow{2}{*}{Metric} & \multicolumn{7}{c|}{Attack} \\ \cline{3-9}
& & NO & Trim & Hist & Rand & MPAF & Zhe. & \textbf{FTI} \\
\hline
\multirow{2}{*}{SD} & MAE & 0.274 & 0.274 & 0.274 & 0.273 & 0.274 & 0.274 & \textbf{72.43} \\
& MSE & 0.092 & 0.092 & 0.092 & 0.092 & 0.092 & 0.092 & \textbf{27.53} \\
\hline
\multirow{2}{*}{IQR} & MAE & 0.274 & 0.275 & 0.275 & 0.274 & 0.265 & 0.273 & \textbf{100.0} \\
& MSE & 0.092 & 0.092 & 0.092 & 0.092 & 0.092 & 0.093 & \textbf{100.0} \\
\hline
\multirow{2}{*}{Z-scores} & MAE & 0.274 & 0.274 & 0.274 & 0.274 & 0.275 & 1.102 & \textbf{100.0} \\
& MSE & 0.092 & 0.092 & 0.093 & 0.092 & 0.092 & 0.416 & \textbf{100.0} \\
\hline
\multirow{2}{*}{SVM} & MAE & 0.274 & 100.0 & 100.0 & 0.275 & 100.0 & 0.768 & \textbf{100.0} \\
& MSE & 0.092 & 100.0 & 100.0 & 0.092 & 100.0 & 0.290 & \textbf{100.0} \\
\hline
\end{tabular}
\end{table}

\subsubsection{Evaluations on the Impact of NDT Density}

Given a 20\% proportion of fake NDTs, Figures 6(a)-(d) compare the Median and GLID rules with varying densities of NDTs in the network scenario. The total number of NDTs does not significantly impact the performance of any attack and defense mechanisms, especially for our FTI and GLID strategies, which is consistent with traditional federated learning settings~\cite{zhang2024poisoning}. Under Median aggregation, FTI consistently shows maximal errors, with MAE and MSE exceeding 100 across different NDT densities, indicating the failure of the defense.
This consistency of performance across varying participants in the distributed NDT system suggests that the total number of NDTs does not substantially influence the effectiveness of the attack and defense strategies.

\begin{figure*}
    \centering
    \begin{tabular}{cc}
        \includegraphics[width=0.47\linewidth]{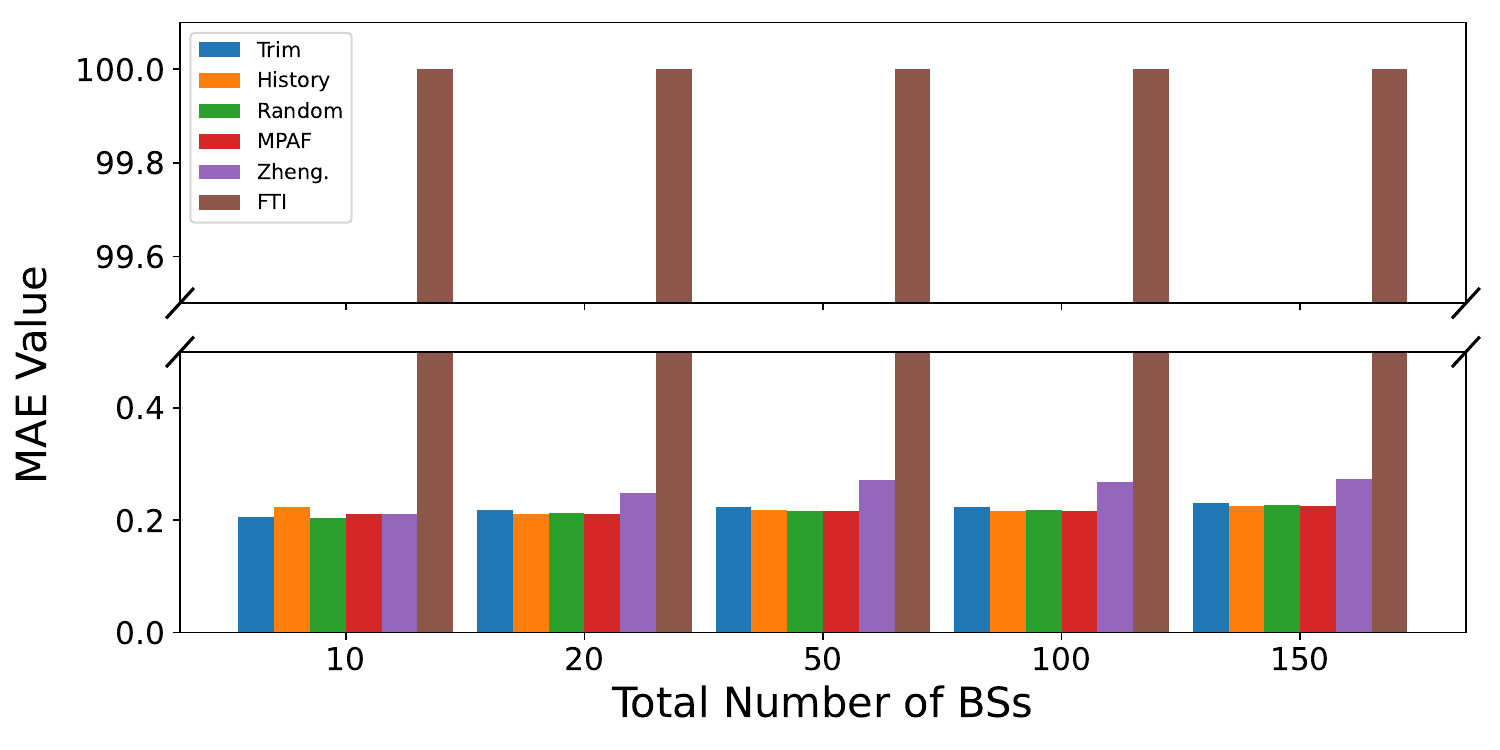} &
        \includegraphics[width=0.47\linewidth]{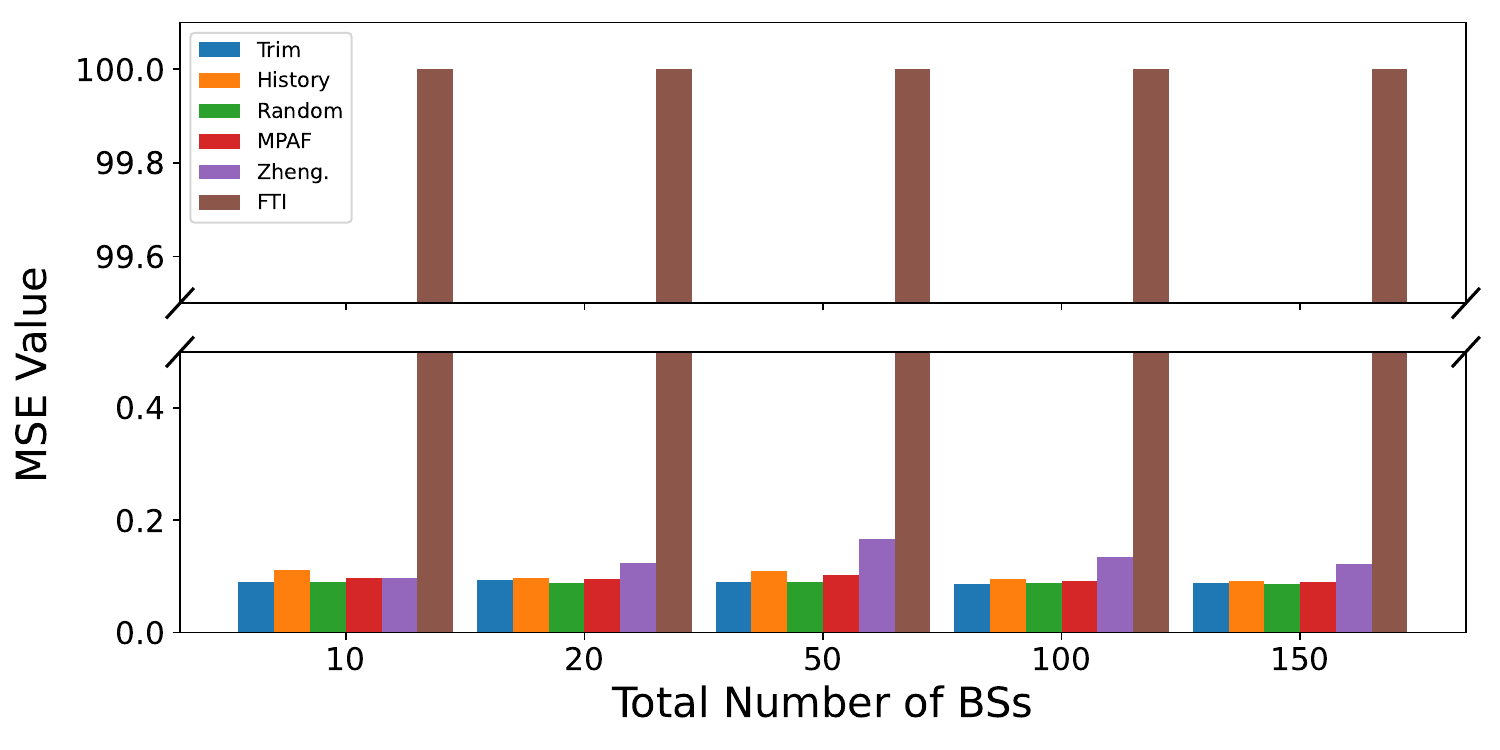} \\
        (a) Median AR w.r.t MAE & (b) Median AR w.r.t MSE \\
        \includegraphics[width=0.47\linewidth]{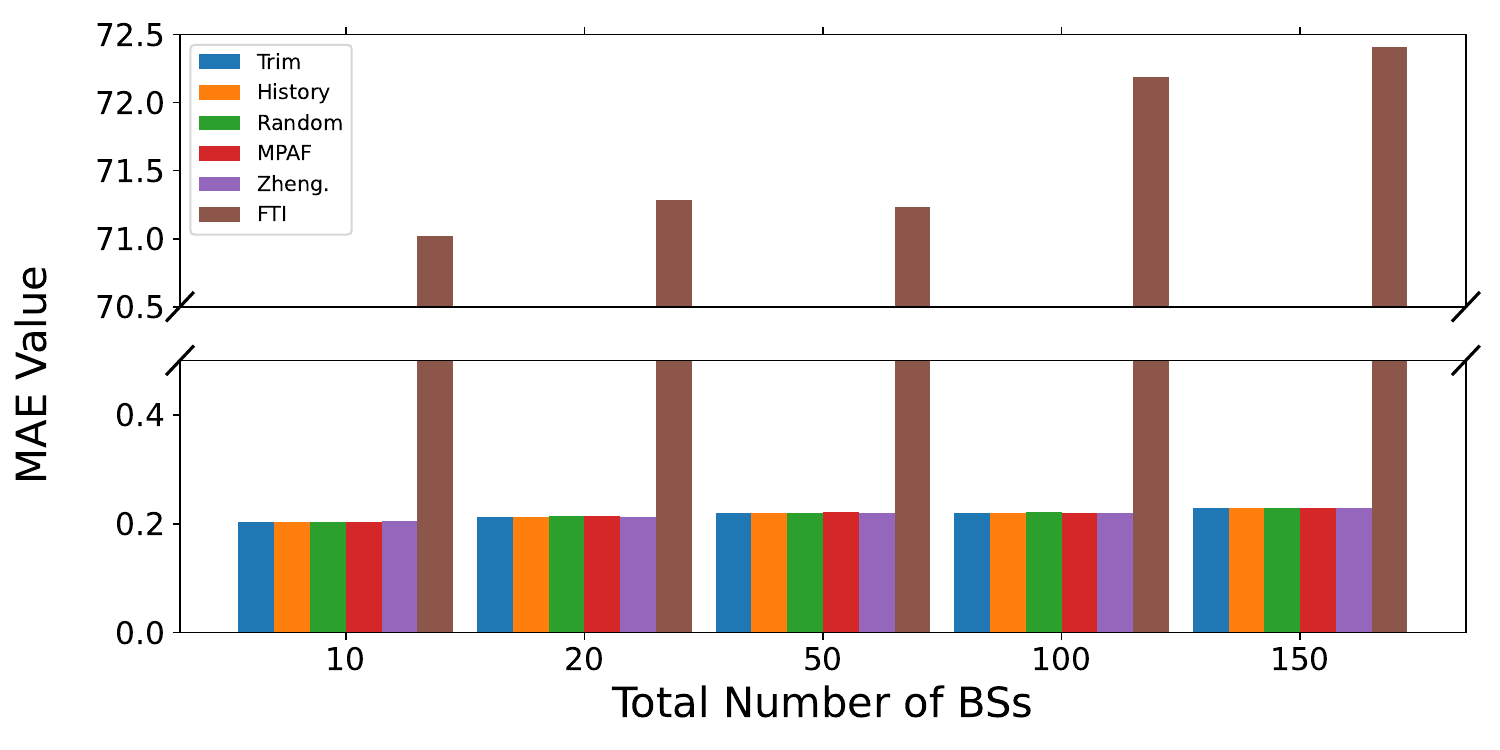} &
        \includegraphics[width=0.47\linewidth]{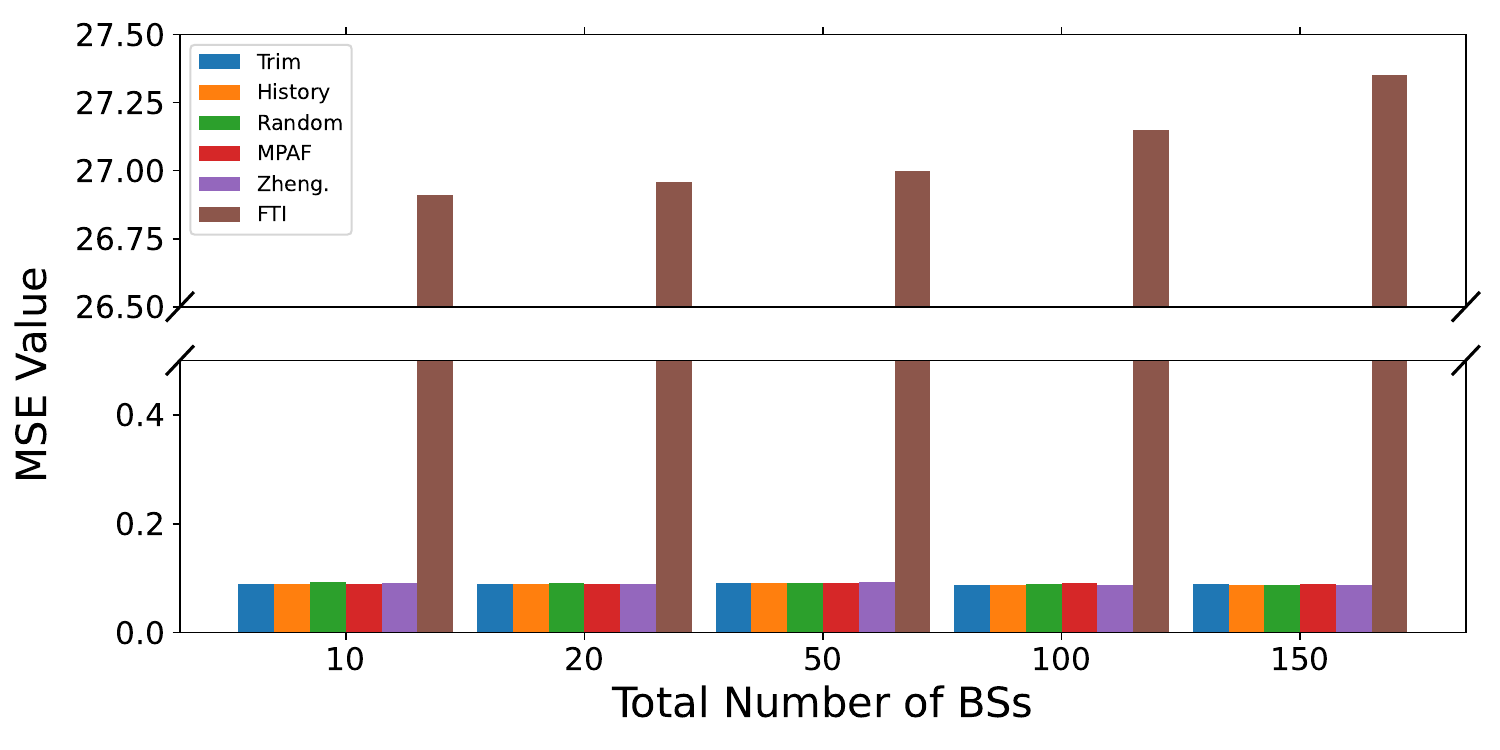} \\
        (c) GLID AR w.r.t MAE & (d) GLID AR w.r.t MSE \\
    \end{tabular}
    \caption{The impact of NDT density on the performance of Median and GLID methods with respect to MAE and MSEs.}
    \label{fig:combined_figures}
    \vspace{-0.5cm}
\end{figure*}

\subsubsection{Evaluations on the Percentile Range of GLID}
\label{sec:percentile}

\begin{table}[ht]
\centering
\caption{Impact of Different Percentile Pairs}
\label{tab:percentile}
\scriptsize
\begin{tabular}{|c|c|c|c|c|c|c|c|}
\hline
\multirow{2}{*}{Pair} & \multirow{2}{*}{Metric} & \multicolumn{6}{c|}{Method} \\ \cline{3-8}
                      &                         & Trim & Hist & Rand & MPAF & Zhe. & \textbf{FTI} \\ \hline
\multirow{2}{*}{[10, 70]} & MAE                  & 100.0 & 100.0 & 100.0 & 100.0 & 0.880 & \textbf{100.0} \\
& MSE                  & 100.0 & 100.0 & 100.0 & 100.0 & 0.346 & \textbf{100.0} \\
\hline
\multirow{2}{*}{[20, 70]} & MAE                  & 0.266 & 0.265 & 0.269 & 0.268 & 0.267 & \textbf{100.0} \\
& MSE                  & 0.102 & 0.106 & 0.104 & 0.101 & 0.106 & \textbf{100.0} \\
\hline
\multirow{2}{*}{[30, 70]} & MAE                  & 0.269 & 0.271 & 0.272 & 0.266 & 0.268 & \textbf{79.634} \\
& MSE                  & 0.112 & 0.109 & 0.110 & 0.106 & 0.109 & \textbf{29.849} \\
\hline
\multirow{2}{*}{[10, 80]} & MAE                  & 100.0 & 100.0 & 100.0 & 100.0 & 0.883 & \textbf{100.0} \\
& MSE                  & 100.0 & 100.0 & 100.0 & 100.0 & 0.340 & \textbf{100.0} \\
\hline
\multirow{2}{*}{[20, 80]} & MAE                  & 0.268 & 0.266 & 0.269 & 0.265 & 0.267 & \textbf{76.468} \\
& MSE                  & 0.106 & 0.102 & 0.104 & 0.101 & 0.106 & \textbf{28.776} \\
\hline
\multirow{2}{*}{[30, 80]} & MAE                  & 0.271 & 0.269 & 0.271 & 0.267 & 0.268 & \textbf{75.411} \\
& MSE                  & 0.109 & 0.110 & 0.106 & 0.109 & 0.112 & \textbf{28.619} \\
\hline
\multirow{2}{*}{[10, 90]} & MAE                  & 100.0 & 100.0 & 100.0 & 100.0 & 0.884 & \textbf{100.0} \\
& MSE                  & 100.0 & 100.0 & 100.0 & 100.0 & 0.339 & \textbf{100.0} \\
\hline
\multirow{2}{*}{[20, 90]} & MAE                  & 0.266 & 0.268 & 0.269 & 0.267 & 0.265 & \textbf{100.0} \\
& MSE                  & 0.109 & 0.106 & 0.105 & 0.110 & 0.106 & \textbf{100.0} \\
\hline
\multirow{2}{*}{[30, 90]} & MAE                  & 0.268 & 0.269 & 0.271 & 0.267 & 0.266 & \textbf{100.0} \\
& MSE                  & 0.106 & 0.109 & 0.110 & 0.105 & 0.109 & \textbf{100.0} \\
\hline
\end{tabular}
\end{table}

Table~\ref{tab:percentile} presents an evaluation of performance across a variety of percentile pairs used in the proposed GLID method on different attack methods. The configuration of the percentile pair guides the GLID method in identifying and eliminating outliers. For example, specifying a percentile pair of [10, 70] means that values below the 10$^{\rm th}$ percentile and above the 70$^{\rm th}$ percentile are trimmed away, focusing the analysis on the data within these bounds.
It is observed that, when the percentile pair is set at [10, 70], most methods, except for the Zheng Attack, register a metric over 100.0, suggesting the models are fully attacked. Similarly, the percentile pair of [10, 90] yields a value over 100 for all methods except the Zheng Attack. The Zheng attack consistently records low metrics across all settings, such as 0.880 and 0.346 for the pair [10, 70], raising questions about its attack efficacy. On the other hand, the FTI attack shows varied performance; it achieves over 100.0 for most percentile pairs like [10, 70] and [20, 90] but drops to 79.634 and 29.849 for the pair [30, 70].
These results underscore the importance of fine-tuning the percentile pair parameters in the GLID method. Proper parameter selection can effectively trim outliers without significantly impacting overall network performance.

\section{Conclusion and Future Work}
\label{sec:conclusion}
In this study, we introduced a novel approach to perform model poisoning attacks on network digital twins through fake traffic injection. Operating under the assumption that real-world BSs are challenging to attack, we inject fake traffic distribution within NDTs with minimum knowledge that disseminates malicious model parameters into distributed network systems. 
Furthermore, we presented an innovative global-local inconsistency detection mechanism, designed to safeguard the NDT systems. 
It employs an adaptive trimming strategy, relying on percentile estimations that preserve accurate model parameters while effectively removing outliers. 
Extensive evaluations demonstrate the effectiveness of our attack and defense, outperforming existing baselines.

With the advent of the digitalization era, the development of an effective security framework for NDT systems presents numerous opportunities for future research and development. Future work could focus on enhancing the capabilities of secure NDT to incorporate real-time data streams and predictive analytics, enabling proactive security management and optimization. Additionally, exploring the integration of explainable artificial intelligence to elucidate model aggregation decisions, detect biases, and ensure the reliability and trustworthiness of the models is a crucial area of research. Further, investigating the application of secure DTs in emerging technologies such as the IoT and 6G cellular systems offers promising avenues for integrated intelligence and autonomy.

\section*{Acknowledgment}
This work was supported by the National Science Foundation under Award SaTC--2350075, CNS--2312138, CNS--2312139, and SaTC--2350076.


\bibliographystyle{IEEEtran}
\bibliography{refs}


 




\vfill

\end{document}